# Slow Light Waveguides based on Bound States in the Continuum

Yuta Tanimura[1]*, Yuki Ishii[1], Kenta Takata[2], Takahiro Uemura[3], Masaya Notomi[2,3,4], Satoshi Iwamoto[5] and Yasutomo Ota[1]†

[1]Department of Applied Physics and Physico-Informatics, Keio University, Yokohama, Kanagawa 223-8522, Japan
[2]NTT Basic Research Laboratories, NTT Corporation, 3-1 Morinosato-Wakamiya, Atsugi 243-0198, Kanagawa, Japan
[3]Department of Physics, Institute of Science Tokyo, 2-12-1 Ookayama, Meguro, 152-8550, Tokyo, Japan
[4]Nanophotonics Center, NTT Corporation, 3-1 Morinosato-Wakamiya, Atsugi, Kanagawa 243-0198, Japan
[5]Research Center for Advanced Science and Technology, The University of Tokyo, Meguro, Tokyo 153-8904, Japan
*y.t.tanimura@keio.jp, †ota@appi.keio.ac.jp

**The concept of bound states in the continuum (BIC) has been advancing light confinement technology in leaky environments. In this letter, we propose and numerically demonstrate a slow light waveguide based on a BIC mode. We considered a waveguide with a polymer core loaded on a plane slab, which supports a leaky guided mode coupled to the radiation continuum in the slab. We found that periodic modulation of the polymer core along the propagation direction can result in a high group index mode with a low propagation loss due to BIC confinement. The introduction of one-dimensional photonic crystals into the BIC waveguides will largely expand its functionality and applications in integrated photonics.**

Bound states in the continuum (BIC) are localized waves embedded in the radiation continuum and have been employed for advanced light manipulation in photonics[1–3]. Photonic structures, including photonic crystals, metasurfaces and coupled waveguide array, have shown light confinement through BIC, which leads to tight light confinement useful for laser oscillation, spectral filtering and environmental sensing. Besides, simple waveguide structures are known to exhibit light waveguiding with BIC modes, in which low loss waveguide modes are formed by cancelling out radiation from otherwise leaky guided modes[4,5]. An example of such BIC waveguide modes is transverse-magnetic-like (TM-like) modes in shallow-etched ridge waveguides[6–13], which have a low effective refractive index and thus couple to the radiation continuum of transverse-electric-like (TE-like) slab modes after polarization conversion. Such a leaky TM waveguide can turn into a low loss BIC mode with careful tuning of structural parameters to induce destructive interference in the TE radiation channel. A similar phenomenon has also been observed in low effective index TE waveguides coupled to high effective index TE slab modes[14,15].

The BIC waveguide structures are useful for various integrated photonics applications. The shallow-etched ridge structures are known to be effective to reduce optical loss induced by etched side wall[12]. Dissipation-free TM modes will be necessary to achieve polarization diverse operation in the waveguide system. The TM guided modes have also been examined as resonant cavity structures embedded in the continuum modes in the TE slab[10,16]. The BIC modes are also found in etch-less structures, such as a plane dielectric thin slab with a low refractive index polymer waveguide on top[17,18]. This approach enables low loss TM waveguides and high-$Q$ resonators without patterning etching-hard materials such as thin film lithium niobate[18]. Furthermore, controllable dissipation from the BIC modes has been employed for investigating non-Hermitian physics in the integrated photonics platform[19].

Improvement of the optical confinement in the BIC waveguide modes has been of continuous interest for more than a decade. Tailoring waveguide cross-section structures or adding auxiliary structures next to the waveguide core has shown to expand the operation wavelength range or reduce structural sensitivity to loss, which facilitates their deployment in practical applications[11,12,20–22]. However, there has been far less discussion on periodic modulation of the BIC waveguide along the propagation direction, or on the introduction of one-dimensional (1D) photonic crystals (PhCs). Such structural modulation will largely modify their dispersion curves and thus will enable various functionalities in the waveguide system, such as controlled radiation to free space, light reflection by photonic bandgaps and slow light propagation[23].

In this letter, we report slow light waveguides based on a BIC mode found in a polymer-core-loaded plane slab structure by introducing a 1D PhC. We sinusoidally modulated the sidewall of the polymer core structure and numerically studied the properties of TM-like guided modes around the first photonic bandgap. For certain modulation depths and around the gap, a TM mode reduces its group velocity as well as its loss, resulting in a BIC-mode slow light propagation. We explain the observed behavior based on a coupled mode theory. Our findings shed new light on the functionality of the BIC waveguide systems. We note that a BIC waveguide with a PhC in the auxiliary side structures has been reported but without discussing the details of its behavior[21]. A PhC waveguide exhibiting an effective refractive index of zero was also reported using BIC confinement but it cannot function as a slow light waveguide[24].

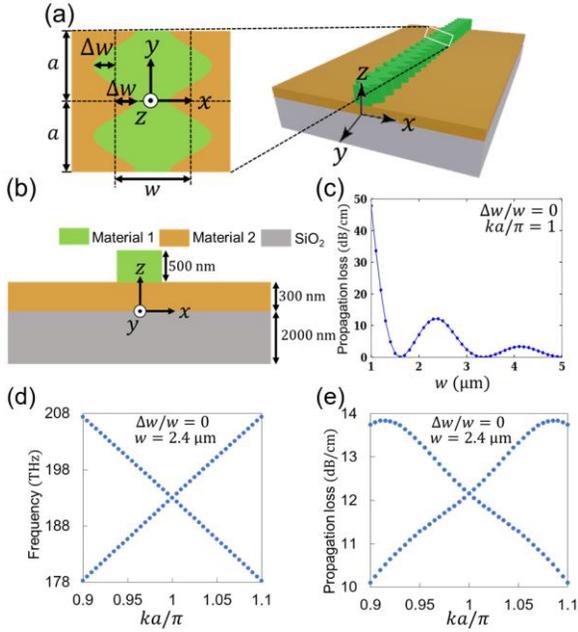

Fig.1. (a) Overview and (b) cross-section of the investigated waveguide structure schematic. The left most inset details the PhC modulation. (c) $w$ dependence of the propagation loss computed for the unmodulated structure with ($\Delta w = 0$). (d) Real and (e) imaginary dispersion curves for the unmodulated waveguide when $w = 2.4$ μm. The numerical results were of the FEM simulations.

Figure 1(a) shows the waveguide structure investigated in this study and Fig. 1(b) depicts its cross-section. A 300-nm-thin slab (material 2) with a relatively high refractive index of $n_2 = 2.36$ is placed on a glass substrate ($n_{SiO_2} = 1.44$). On top of the slab, a low-refractive index polymer core (material 1, $n_1 = 1.54$) with a height of 500 nm is loaded. The polymer ridge is designed to have a base width of $w$ and its sidewalls are modulated sinusoidally with a depth of $\Delta w$, as illustrated in the figure. The period of the modulation was set to $a = 450$ nm. The choices of the structural and material parameters are arbitrary and will not critically affect on the conclusions obtained in this study. We numerically analyzed the structure using the 3D finite element method (FEM) in the telecommunication wavelength bands (COMSOL Multiphysics). The $x$ and $z$ boundaries are terminated with perfectly matched layers and a periodic boundary condition was applied in $y$ direction.

We start the discussion by examining unmodulated waveguide structures by setting $\Delta w = 0$. TM guided modes in the structure can couple to the TE-polarized radiation continuum in the plane slab due to TE-TM conversion and thus are inherently dissipative. Figure 1(c) shows propagation losses calculated for the fundamental TM guided mode as a function of $w$ at a wavelength of 1550 nm. A damping oscillation of loss with respect to $w$ is found. For certain $w$, strong loss suppressions are observed, which correspond to the formation of low loss BIC waveguide modes[17,18]. Indeed, a propagation loss of $7 \times 10^{-4}$ dB/cm is calculated for $w = 1.62$ μm, which suggests nearly-perfect elimination of the radiative coupling from the TM mode to the slab modes by destructive interference. Figures 1(d) and (e) respectively show the real and imaginary parts of the dispersion curve for the case $w = 2.4$ μm and $\Delta w = 0$. Almost linear crossings of the dispersions at X point (where wavenumber $k$ equals $\pi/a$) are visible, which were caused by the fictious zone folding due to the application of the periodic boundary condition. The monotonic change of the imaginary part indicates the absence of the BIC confinement for the case of $w = 2.4$ μm in the wavelength range of interest.

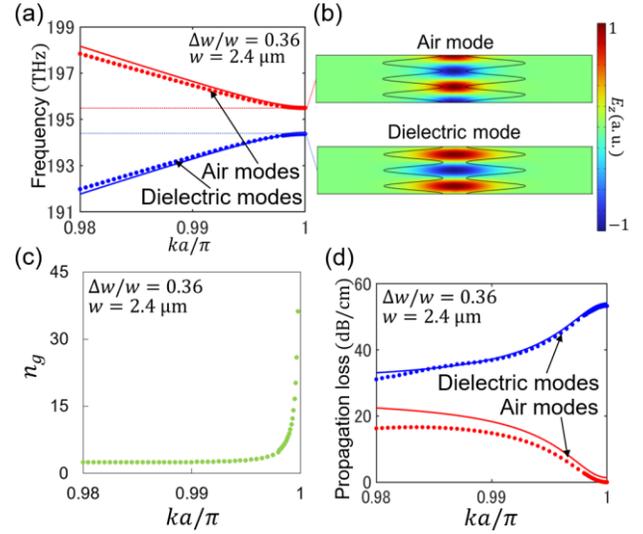

Fig.2. (a) Real-value dispersion curves of the air and dielectric modes computed with $w = 2.4$ μm and $\Delta w = 0.36w$. Solid lines are of the theoretical mode discussed in the main text. (b) $E_z$ field distributions for the investigated two modes evaluated at X point. (c) $n_g$ of the air mode as a function of $k$ and (d) Propagation losses of the two modes.

Next, we investigate the influence of the periodic modulation on the fundamental TM guided mode for the structure with $w = 2.4$ μm. Figure 2(a) shows computed dispersion curves of the 1D PhC around X point. A photonic bandgap of 1.12 THz appears around 195 THz, which corresponds to the first photonic bandgap in the dispersion diagram. Figure 2(b) displays computed eigenmode distributions represented by the major electric field $E_z$. At X point, the mode of the upper band is mostly localized around the neck of the modulated structure, which we term as an air mode. The mode of the lower band distributes in the bulge and is termed as a dielectric mode. Both modes show flat dispersions around the gap, suggesting their propagation as slow light modes. Figure 2(c) shows an evolution of the group index $n_g$ of the air mode, which behaves the same with that of the dielectric mode. While $n_g$ remains only 2.4 in the dominant part of the plot, it shows a significant increase in the vicinity of X point and reaches 7.1 at $k = 0.9987\pi/a$. We note that $n_g$s are evaluated from the dispersion curve using the backward-difference approximation. Then, we evaluated optical losses of the TM guided modes as plotted in Fig. 2(d). The two optical bands exhibit different loss values and their splitting becomes prominent as approaching X point. At $k = 0.9998\pi/a$ where $n_g = 36$, the loss of the air band is only 3.4×10$^{-2}$ dB/cm, demonstrating low-loss slow light propagation in the air band mode. We note that the loss splitting behavior around the symmetric point in the momentum space resembles those observed in second-order distributed feedback (DFB) resonators[25]. However, our case largely differs and focuses on the first order grating resonance below the light line and the loss process is dominated by zeroth order diffraction. The loss splitting phenomenon observed here can be used for laser mode selection in a DFB laser.

To understand how the low-loss slow light mode was formed in the modulated waveguide, we developed a theoretical model based on a coupled mode theory. The dominant loss mechanism for the TM mode, whose electric field vector is hereafter expressed as $\boldsymbol{E}^{\text{TM}}(\boldsymbol{r})$, is its coupling to the unbound TE slab modes ($\boldsymbol{E}^{\text{TE}}(\boldsymbol{r})$), where $\boldsymbol{r}$ is the spatial coordinate. Thus, evaluating mutual coupling strength is essential to account for the TM mode loss. The coupling constant between the two modes as the zeroth order diffraction can be described by[26,27]

$$\kappa \propto \int_{period} \left( \iint \Delta \epsilon \boldsymbol{E}^{\text{TM}*} \cdot \boldsymbol{E}^{\text{TE}} dxdz \right) dy \quad (1)$$

, where the integral is taken over the entire unit cell of the PhC. $\Delta\epsilon$ expresses permittivity perturbation induced by the loading of the polymer waveguide on the slab and is therefore nonzero only where the polymer is present. Since unperturbed $\boldsymbol{E}^{\text{TE}}$ is dominantly polarized in the $xy$ plane, only in-plane electric field components of $\boldsymbol{E}^{\text{TM}}$ can contribute to the inner product. We found that $E_x^{\text{TM}}$ is about an order of magnitude weaker than $E_y^{\text{TM}}$. Thus we safely ignored its contribution to the integral. Figure 3(a) shows in-plane electric field ($E_y^{\text{TM}}$) distributions of the air and dielectric TM modes calculated at X point. $E_y$ field of the dielectric mode is localized in the neck of the waveguide, rather than in the bulge, as opposed to the case of the $E_z$ field. The $E_y$ field is well localized around a specific $y$ position in the polymer so that its contribution to $y$ integral may be represented by the value at the localized position. Moreover, the $z$ integral in Eq (1) can be assumed to provide a constant contribution determined solely by the vertical confinement of the modes. With the above assumptions, the equation of $\kappa$ reduces to

$$\kappa \propto \int \Delta\epsilon E_y^{\text{TM}*} \cdot E_y^{\text{TE}} dx. \quad (2)$$

To analytically calculate eq. (2), we assume $\boldsymbol{E}^{\text{TE}}$ as a guided plane wave with a propagation constant of $n_{\text{TE}}k_0$, where $n_{\text{TE}}$ is the effective refractive index of the TE mode in the slab and $k_0$ is the wavenumber in air. To induce the TE-TM mode coupling, the two modes need to be phase matched in the $y$ direction, posing that both modes share a propagation constant in the $y$ direction of $n_{\text{TM}}k_0$. Considering the momentum conservation, the wavefunction of the TE mode propagating in $x$ direction can be expressed as

$$E_y^{\text{TE}} \propto \exp\left(i\sqrt{n_{\text{TE}}^2 - n_{\text{TM}}^2}k_0 x\right) \quad (3)$$

To further simplify the model, we assumed that the $x$ profile of $E_y^{\text{TM}}$ is constant. Then, the integral can be analytically performed and the optical loss, which is proportional to $|\kappa|^2$[27], can be derived for the air ($|\kappa_{\text{air}}|^2$) and dielectric ($|\kappa_{\text{die}}|^2$) modes as follows:

$$|\kappa_{\text{air}}|^2 \propto \gamma_{\text{air}}^2 \left[1 - \cos\left\{\sqrt{n_{\text{TE}}^2 - n_{\text{TM}}^2}k_0(w + \Delta w)\right\}\right] \quad (4a)$$

$$|\kappa_{\text{die}}|^2 \propto \gamma_{\text{die}}^2 \left[1 - \cos\left\{\sqrt{n_{\text{TE}}^2 - n_{\text{TM}}^2}k_0(w - 2\Delta w)\right\}\right] \quad (4b)$$

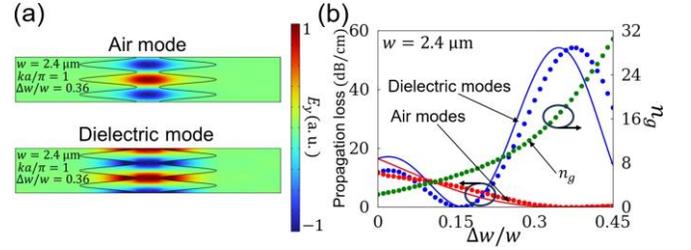

Fig.3. (a) $E_y$ field distributions for the two modes calculated at X point for the structure with $w = 2.4$ μm and $\Delta w/w = 0.36$. (b) $\Delta w$ dependence of propagation loss at X point and $n_g$ at $k = 0.999\pi/a$ computed for the waveguide with $w = 2.4$ μm. The solid lines are of the theoretical model.

, where $\gamma_{\text{air}}$ and $\gamma_{\text{die}}$ are introduced as empirical $\kappa$ increase and suppression factors, respectively and are formulated as

$$\gamma_{\text{air}} = \gamma_0 w_0 (w_0 + \Delta w)^{-1} \quad (5a)$$
$$\gamma_{\text{die}} = \gamma_0 (w_0 + \Delta w)/w_0 \quad (5b)$$

These factors suggest that the air (dielectric) mode becomes less (more) likely to couple to the TE continuum with increasing $\Delta w$. $\gamma_0$ is a normalization factor and $w_0$ is set to be 1 μm. We note that, in the derivation of eq(4a) and (4b), we postulated that $E_y^{\text{TM}}$ becomes zero when $|x| > (w + \Delta w)/2$ to reflect the weak field amplitude around the edge of the polymer.

The derived model suggests that loss will oscillate with $\Delta w$ and will vanish for certain choices of parameters, which corresponds to the BIC optical confinement. To validate the analytical model, we compare its prediction with numerical simulations as summarized in Fig. 3(b), which plots losses of the air and dielectric modes at X point as a function of $\Delta w$. In the analytical calculation (solid lines), we used the $n_{\text{TE}}$ and $n_{\text{TM}}$ values for the unperturbed structures evaluated by separate numerical computations. We also adjusted $\gamma_0$ to align the peak loss values of the numerical and analytical results. As expected, the numerically computed losses (red and blue dots) show oscillatory behaviors and take minimam at certain $\Delta w$ values. The lowest losses are 1.6×10$^{-2}$ dB/cm for the dielectric mode at $\Delta w/w = 0.16$ and 9.8×10$^{-4}$ dB/cm for the air mode at $\Delta w/w = 0.36$. The reasonable agreement of the analytical model with the numerical simulations suggests that the observed loss suppression arises from the cancellation of the coupling to the TE radiation continuum, namely BIC. Figure 3(b) shows computed $n_g$ values evaluated at $k = 0.999\pi/a$ and plotted as a function of $\Delta w$. We observed a monotonic increase of $n_g$ with $\Delta w$. The increase of $n_g$ can be associated with the enhancement of photonic bandgap, which is primarily determined by the potential difference between the neck and bulge region of the PhC. A wider gap bends the dispersion more strongly, leading to larger $n_g$. The dispersion curves of the air and dielectric bands can also be analytically modeled using a non-Hermitian coupled mode theory[28]. We assume that light at X point is tightly localized either at the neck or bulge of the waveguide and the photons in the two leaky cites are coherently and dissipatively coupled each other. Under this assumption, non-Hermitian dispersion curves can be analytically calculated with a Bloch boundary condition. The analytical results are overlaid in Figs. 2(a) and (d) and reasonably reproduce the numerically obtained dispersion curves.

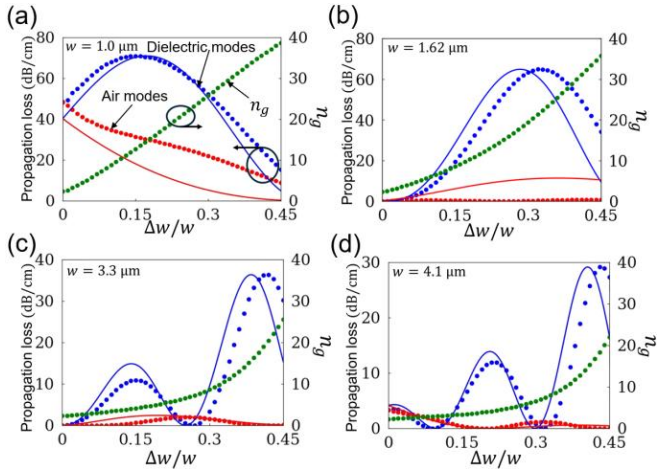

Fig.4. Propagation losses at X point and $n_g$ as a function of $\Delta w$ for the structures with (a) $w = 1.0$ μm, (b) $w = 1.62$ μm, (c) $w = 3.3$ μm, and (d) $w = 4.1$ μm. $n_g$ values were evaluated at $k = 0.999\pi/a$ for the air mode. Solid lines are of the theoretical model.

Finally, we discuss the influence of $w$ on the BIC slow light propagation. The dot plots in Fig. 4 show $\Delta w$ dependences of loss and $n_g$ computed for different $w$s. The solid lines are from the analytical model, which takes into account the $w$ dependence of $n_{TM}$ of the unperturbed waveguide. For $w = 1$ μm displayed in Fig. 4(a), no BIC condition was numerically found for all the possible $\Delta w$. In this design, overall loss is high for both the air and dielectric modes. For $w = 1.62$ μm, which corresponds to the first BIC condition for the unmodulated structure (see Fig. 1(c)), a strong suppression of loss for the air mode over a wide range of $\Delta w$ was found. At the lowest loss condition at $\Delta w/w = 0.23$, a loss of $2.4 \times 10^{-3}$ dB/cm with $n_g = 14$ is achieved. In this case, the analytical model does not explain this observation. We assume that the oversimplification of the model is a reason and indeed found a better reproduction of the observation by considering more details of the relevant optical modes (not shown). However, the more complex model does not comprehensively explain the behaviors of the structures with different $w$. The remaining inconsistency suggests more room for theoretical study, which we will address in future work. Setting $w = 3.3$ μm is the second BIC condition of the unperturbed waveguide and shows a reasonable matching with the theory. We noticed the overall loss becomes lower by increasing $w$, which is probably due to tighter optical confinement in the polymer core and the resulting suppression of the TE-TM conversion. At $\Delta w/w = 0.43$, the air mode shows a loss of $5.5 \times 10^{-3}$ dB/cm at X point with $n_g = 22$ at $k = 0.999\pi/a$. The value of $n_g$ is lower compared to those calculated for the structure with smaller $w$ at the same $\Delta w$. This suggests that the effect of the modulation becomes weaker as increasing $w$. The same trend is confirmed for the case with $w = 4.1$ μm (Fig. 4(d)). Overall, the loss behaviors for different structures are reasonably reproduced by the single theoretical model, further confirming the BIC optical confinement in the investigated slow light waveguides.

In conclusion, we demonstrated a BIC slow light waveguide by introducing a 1D PhC into a polymer-loaded dielectric slab structure. Provided a proper depth of the PhC modulation, the waveguide simultaneously exhibited a high $n_g$ and a low propagation loss. The loss suppression mechanism was found to be due to the BIC optical confinement through fitting to a theoretical model. We foresee that the same loss suppression approach can be widely applicable to various lossy 1D PhCs. The BIC slow light waveguide does not require nanopatterning of the slab material itself but exhibits largely modified dispersions and can enhance light matter interactions, which will expand the functionality and applications of the BIC waveguides.

**Funding.** Fusion Oriented Research for disruptive Science and Technology (JPMJFR213F), JST CREST (JPMJCR19T1), Japan Society for the Promotion of Science (22H00298, 22H01994, 22K18989), Iketani Foundation, Nippon Sheet Glass Foundation.

**Acknowledgment.** We gratefully acknowledge Prof. T. Ozawa for insightful discussion.

**Disclosures.** The authors declare no conflicts of interest.

**Data availability.** Data underlying the results presented in this paper are not publicly available at this time but may be obtained from the authors upon reasonable request.